\documentclass[10pt,twocolumn,letterpaper]{article}
\usepackage[letterpaper,top=1in,bottom=1.125in,left=0.8125in,right=0.8125in,columnsep=0.3125in]{geometry}
\usepackage{times}
\usepackage[utf8]{inputenc}
\usepackage[T1]{fontenc}
\usepackage{amsmath,amssymb,amsfonts}
\usepackage{graphicx}
\usepackage{booktabs}
\usepackage{multirow}
\usepackage{nicefrac}
\usepackage[dvipsnames]{xcolor}
\usepackage{microtype}
\usepackage[numbers,sort&compress]{natbib}
\usepackage[format=plain,labelformat=simple,labelsep=period,font=small]{caption}
\usepackage[font=footnotesize]{subcaption}
\usepackage{cuted}
\usepackage[hyphens]{url}
\usepackage[shortlabels,inline]{enumitem}
\definecolor{linkblue}{rgb}{0.21,0.49,0.74}
\usepackage[breaklinks,colorlinks,allcolors=linkblue]{hyperref}
\newcommand{\best}[1]{\textcolor{red}{\textbf{#1}}}
\newcommand{\second}[1]{\textcolor{blue}{\underline{#1}}}
\title{FDIR: Fidelity-Anchored Decompressed Image Restoration for Human and Machine Perception}
\author{%
Kuan-Yen Chen\thanks{Equal contribution.} \\ National Cheng Kung University \\ {\small\ttfamily azure0413@iir.csie.ncku.edu.tw}
\and Chung-Yao Ma\footnotemark[1] \\ National Taiwan University \\ {\small\ttfamily d13922024@csie.ntu.edu.tw}
\and Fang-Yi Su\footnotemark[1] \\ Academia Sinica \\ {\small\ttfamily fangyisu994@as.edu.tw}
\and Philip Chikontwe \\ \makebox[0.4\textwidth]{Harvard Medical School} \\ {\small\ttfamily philip\_chikontwe@hms.harvard.edu}
\and Jonathan Guo \\ \makebox[0.4\textwidth]{Independent Researcher} \\ {\small\ttfamily ougnahtanoj@gmail.com}
\and Pu-Jen Cheng \\ \makebox[0.4\textwidth]{National Taiwan University} \\ {\small\ttfamily pjcheng@csie.ntu.edu.tw}
\and Jung-Hsien Chiang \\ \makebox[0.4\textwidth]{National Cheng Kung University} \\ {\small\ttfamily jchiang@mail.ncku.edu.tw}
}
\date{}
\begin{document}
\maketitle
\begin{strip}
\centering
\includegraphics[width=\linewidth]{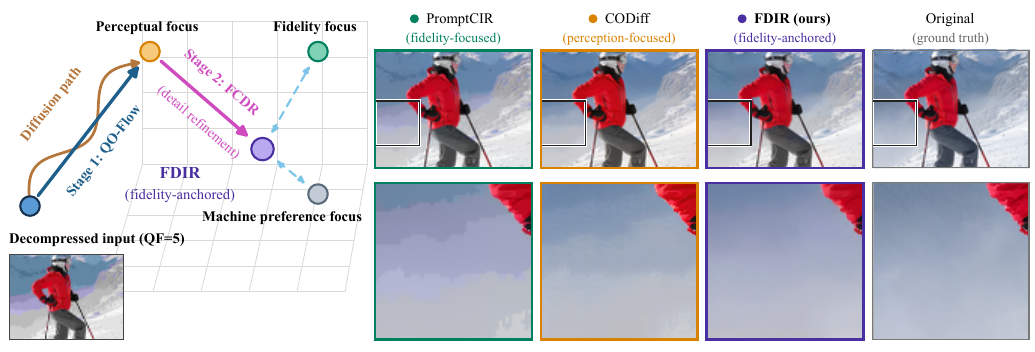}
\captionsetup{hypcap=false}
\captionof{figure}{\textbf{Overview.} Existing methods tend to favor one criterion. In FDIR, QO-Flow generates a structural prior in one step, and FCDR uses this prior to predict a residual on the decompressed input. Right: results on an input with quality factor (QF) 5, with a zoomed region below. Frame colors match the nodes on the left.}
\captionsetup{hypcap=true}
\label{fig:teaser}%
\end{strip}

\begin{abstract}
Evaluating restored images requires considering pixel fidelity, human perception, and machine preference. However, existing compressed image restoration methods rarely satisfy all three criteria at once. Regression models preserve pixel fidelity but often produce overly smooth results. Conversely, generative models create realistic textures but tend to drift from the ground truth. To address this trade-off, we propose FDIR (Fidelity-Anchored Decompressed Image Restoration), a framework that uses generative priors while preserving strong pixel fidelity. FDIR works in two stages. First, a quality-guided one-step flow model (QO-Flow) adapts a pretrained latent flow backbone with LoRA to generate a structural prior in a single forward pass. Second, rather than outputting this generated image directly, a flow-conditioned detail refinement network (FCDR) uses it to predict a pixel-space residual on the decompressed input. This design brings in perceptual details while keeping the final output faithful to the original image. We evaluate FDIR on LIVE1 and DIV2K under severe compression at quality factors 1 and 5. Across all four settings, FDIR achieves the highest PSNR among the seven compared methods and beats the strongest perceptual baseline, CODiff, by 1.3 to 2.1 dB. It also ranks second on LPIPS in all four settings and on FID and DISTS on DIV2K-Val, right behind CODiff. For downstream detection, segmentation, and retrieval at quality factors 20 and 30, FDIR ranks second on every individual metric and achieves the best average rank across all ten metrics. Overall, FDIR is the only method to maintain a top-two average rank across fidelity, perception, and machine preference.
\end{abstract}

\section{Introduction}
\label{sec:intro}

Image restoration (IR) recovers high-quality images from degraded inputs \citep{buadesReviewImageDenoising2005, suSurveyDeepLearning2022}. Most methods are designed for and judged by a single criterion, either pixel fidelity or perceptual quality~\citep{blau2018PIRMChallenge2018, guPIPALLargeScale2020}. A restored image, however, is often viewed by humans and processed by vision models. We therefore evaluate restoration on three criteria: pixel fidelity, human perception, and machine preference. Machine preference measures how well vision models perform on the restored image~\citep{liImageQualityAssessment2025}. Few existing methods are assessed on all three~\citep{aiDreamClear2024}.

We focus on compressed image restoration, where the decompressed input from a lossy codec still carries compression artifacts. Codecs such as JPEG~\citep{wallaceJPEGStillPicture1991}, WebP~\citep{zernWebPImageFormat2024}, and AVIF~\citep{aomAV1ImageFileFormat2025} are widely used for digital storage and transmission, so artifact removal has lasting practical value. Unlike linear degradations with a known forward operator~\citep{martinPnPFlowPlugandPlayImage2025}, JPEG compression artifacts come from coarse quantization of transform coefficients \citep{dongCompressionArtifactsReduction2015, hanJDECJPEGDecoding2024}. The resulting blocking and ringing artifacts~\citep{dongCompressionArtifactsReduction2015} call for strong learned priors~\citep{jiangFlexibleBlindJPEG2021}.

Existing approaches face a dilemma. Fidelity-oriented regression methods \citep{dongCompressionArtifactsReduction2015, zhangGaussianDenoiserResidual2017, jiangFlexibleBlindJPEG2021, liangSwinIRImageRestoration2021, liPromptCIRBlindCompressed2024, hanJDECJPEGDecoding2024} prioritize pixel fidelity, but they can produce overly smooth outputs \citep{blauPerceptionDistortionTradeoff2018, freirich2021theory, ledigPhotoRealisticSingleImage2017}. Generative methods \citep{linDiffBIRBlindImage2024a, guoCompressionAwareOneStepDiffusion2025} improve perceptual quality, but the content they synthesize can depart from the ground truth. \citet{cohenLooksTooGood2024} argue that such hallucination is an inherent cost of approaching perfect perceptual quality. Recent work \citep{liImageQualityAssessment2025} further reports that neither fidelity nor perceptual scores reliably predict machine preference.

Our experiments reflect this dilemma. Regression methods such as FBCNN~\citep{jiangFlexibleBlindJPEG2021} rank high on peak signal-to-noise ratio (PSNR) and SSIM~\citep{wangImageQualityAssessment2004} but blur fine textures. The generative baseline CODiff~\citep{guoCompressionAwareOneStepDiffusion2025} leads all perceptual metrics, but under severe compression its PSNR is lower than ours in every setting. This observation motivates a design that uses a generative model as guidance rather than as the final output (Figure~\ref{fig:teaser}).

We propose FDIR, a fidelity-anchored framework for decompressed image restoration. FDIR works in two stages (Figure~\ref{fig:method_overview}). A quality-guided one-step flow model (QO-Flow) adapts a pretrained latent flow backbone~\citep{esserScalingRectifiedFlow2024, lipmanFlowMatchingGenerative2023} with Low-Rank Adaptation (LoRA)~\citep{huLoRALowRankAdaptation2021} and generates a structural prior in a single forward pass. A flow-conditioned detail refinement network (FCDR) then combines this prior with the decompressed input to predict a pixel-space residual added to the input. We call FDIR fidelity-anchored because its output is the decompressed input plus a learned correction, and its objective is dominated by ground-truth reconstruction. Perception and machine preference are improved on top of this anchor. Figure~\ref{fig:teaser} shows an example at quality factor (QF) 5. The fidelity-focused PromptCIR~\citep{liPromptCIRBlindCompressed2024} leaves banding in the smooth background, and the perception-focused CODiff removes it with a lower PSNR. FDIR removes the banding and has the highest PSNR of the three on this image. The resulting model is balanced across metrics. Across the 16 fidelity and perceptual columns of Tables~\ref{tab:extreme_compression} and~\ref{tab:fid_comparison}, FDIR has the best average rank (1.75, compared with 2.75 for CODiff and 2.81 for FBCNN). It never ranks below third in any column, whereas every other method ranks fifth or lower in at least one. Our main contributions are:
\begin{itemize}[leftmargin=1.5em, itemsep=2pt, topsep=2pt]
\item \textbf{A fidelity-anchored two-stage design.} QO-Flow generates a structural prior in one step. FCDR uses this prior to predict a residual on the decompressed input instead of outputting the generated image. Compared with the Stage~1 output, this raises PSNR by 0.80 to 1.52~dB at QF=1, 5, and 10 while LPIPS~\citep{zhangUnreasonableEffectivenessDeep2018} changes by at most 0.012.
\item \textbf{Efficient adaptation of a flow prior.} QO-Flow adapts a pretrained latent flow backbone with LoRA and QF conditioning and requires a single function evaluation. Across QO-Flow and FCDR, only 73.5M parameters are trained, while the 2.1B-parameter backbone remains frozen.
\item \textbf{Balanced results across criteria.} FDIR has the best average rank across the fidelity and perceptual metrics and never ranks below third on any of them (Figure~\ref{fig:radar}). Under severe compression on LIVE1~\citep{sheikhLiveImageQuality2005a} and DIV2K-Val~\citep{agustssonNTIRE2017Challenge2017}, FDIR attains the highest PSNR in all four settings, 1.3 to 2.1~dB above CODiff. It ranks second on LPIPS on both datasets and on FID~\citep{heuselFID} and DISTS~\citep{dingImageQualityAssessment2020} on DIV2K-Val. On ten downstream metrics for detection, segmentation, and retrieval, it ranks second on every metric and has the best average rank. It is the only compared method whose average rank is within the top two on fidelity, perception, and machine preference.
\end{itemize}

\section{Related Work}
\label{sec:related}

\textbf{Restoration of compressed images.}
Deep learning approaches to compression artifact removal have progressed from convolutional neural networks (CNNs) to more expressive architectures. ARCNN~\citep{dongCompressionArtifactsReduction2015} was an early convolutional network for JPEG artifact reduction, and DnCNN~\citep{zhangGaussianDenoiserResidual2017} adopted residual learning~\citep{heDeepResidualLearning2015}, which became a standard choice. FBCNN~\citep{jiangFlexibleBlindJPEG2021} adds quality factor estimation for blind restoration. SwinIR~\citep{liangSwinIRImageRestoration2021} and PromptCIR~\citep{liPromptCIRBlindCompressed2024} use Transformer~\citep{vaswaniAttentionAllYou2017} backbones, and JDEC~\citep{hanJDECJPEGDecoding2024} operates directly on discrete cosine transform (DCT) coefficients. These regression methods can produce overly smooth outputs that lack fine detail~\citep{blauPerceptionDistortionTradeoff2018, ledigPhotoRealisticSingleImage2017}.

\textbf{Generative priors for restoration.}
Diffusion models~\citep{hoDenoisingDiffusionProbabilistic2020} are strong generative priors for IR. SR3~\citep{sahariaImageSuperResolutionIterative2021} demonstrated iterative refinement for super-resolution. DiffBIR~\citep{linDiffBIRBlindImage2024a} uses a pretrained Stable Diffusion prior~\citep{rombachHighResolutionImageSynthesis2022} in a two-stage pipeline that first reduces degradation and then generates details. SUPIR~\citep{yuScalingExcellencePracticing2024} adopts Stable Diffusion XL~\citep{podellSDXL2024} as its prior. Flow Matching (FM)~\citep{lipmanFlowMatchingGenerative2023} learns deterministic ordinary differential equation (ODE) trajectories between distributions, and SD3~\citep{esserScalingRectifiedFlow2024} scales rectified flows~\citep{liuFlowStraightFast2023} to large latent models. FM-based restoration and enhancement methods include PnP-Flow~\citep{martinPnPFlowPlugandPlayImage2025}, FlowIE~\citep{zhuFlowIEEfficientImage2024}, and ResFlow~\citep{qinReversingFlowImage2025}. CODiff~\citep{guoCompressionAwareOneStepDiffusion2025} adapts a diffusion prior for one-step restoration of compressed images. Task-oriented methods~\citep{yangVisualRecognitionDrivenImage2023, chenUniRestoreUnifiedPerceptual2025} instead optimize restoration for downstream recognition. Perception-oriented generative methods can synthesize content that departs from the ground truth, lowering pixel fidelity~\citep{cohenLooksTooGood2024}. In pipelines such as DiffBIR and SUPIR, the generative model produces the final output. By contrast, FDIR uses the generated image only as guidance and outputs a residual correction on the decompressed input. Concurrently, \citet{satohFidelityConstrainedAnchoring2026} anchors the output of a black-box denoiser to its input through training-free blending under a PSNR or SSIM constraint between output and input. In that work, fidelity refers to similarity to the input. FDIR instead learns a pixel-space residual with ground-truth supervision, and fidelity here means similarity to the ground truth.

\textbf{Quality criteria and their trade-offs.}
The perception-distortion trade-off~\citep{blauPerceptionDistortionTradeoff2018, freirich2021theory} states that distortion and the divergence from the natural-image distribution cannot both be minimized. \citet{liuClassificationDistortionPerception2019} extend this trade-off to classification accuracy. \citet{ohayonPerceptionRobustnessTradeoff2024} show that a deterministic estimator requires an unbounded Lipschitz constant to achieve both perfect perceptual quality and perfect consistency with the measurements. Empirically, \citet{liImageQualityAssessment2025} find that neither fidelity nor perceptual quality reliably predicts downstream task performance. FDIR is a deterministic estimator that puts fidelity first, so it operates away from the perfect-perception limit described in these works. We evaluate it on all three criteria.

\section{Method}
\label{sec:method}
\begin{figure*}[t]
\centering
\includegraphics[width=\textwidth]{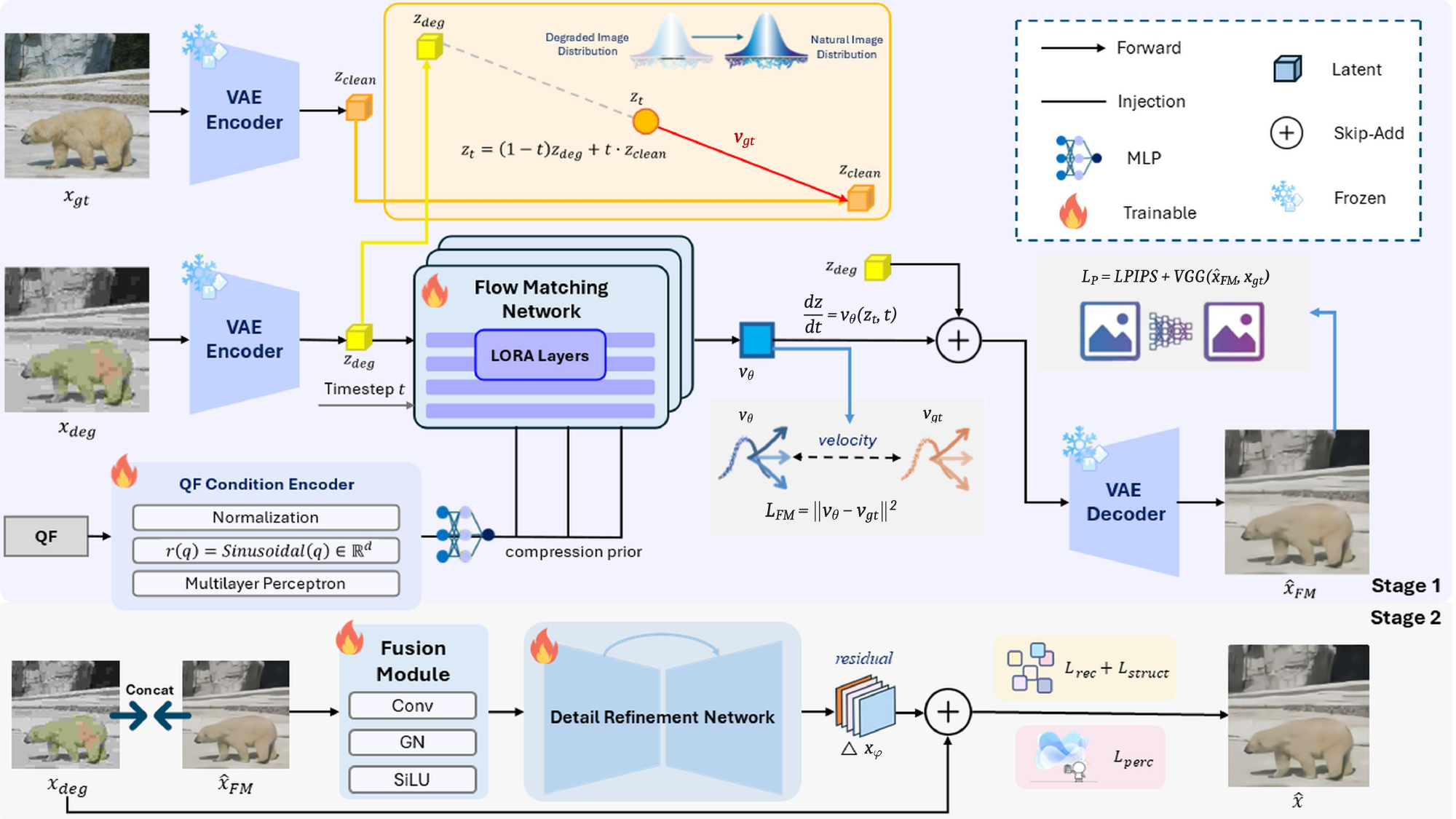}
\caption{\textbf{Overview of FDIR.} Stage~1 (QO-Flow) generates a structural prior in latent space with one step. Stage~2 (FCDR) uses this prior to predict a pixel-space residual on the decompressed input.}
\label{fig:method_overview}
\end{figure*}

\begin{figure}[t]
\centering
\includegraphics[width=0.82\columnwidth]{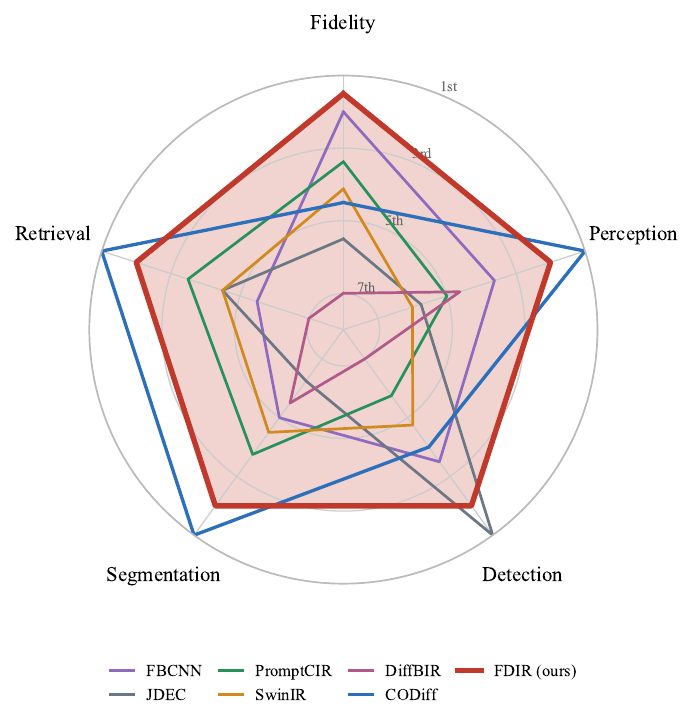}
\caption{\textbf{Mean rank on five axes} among the seven methods (outer is better). Fidelity: PSNR and SSIM in Table~\ref{tab:extreme_compression}. Perception: LPIPS, FID, and DISTS in Tables~\ref{tab:extreme_compression} and~\ref{tab:fid_comparison}. Detection, Segmentation, and Retrieval: the corresponding task metrics in Table~\ref{tab:downstream_tasks}.}
\label{fig:radar}
\end{figure}

FDIR splits restoration into two stages with different roles (Figure~\ref{fig:method_overview}). Stage~1 generates a structural prior in latent space. Stage~2 predicts a pixel-space residual on the decompressed input, guided by this prior.

\subsection{Quality-Guided One-Step Flow Matching (Stage~1)}

JPEG compression coarsely quantizes DCT coefficients and sets many high-frequency coefficients to zero~\citep{wallaceJPEGStillPicture1991}. Recovering the lost content requires a prior on natural images. We adopt Flow Matching~\citep{lipmanFlowMatchingGenerative2023} with a linear path~\citep{liuFlowStraightFast2023} between the degraded and clean latents. The target velocity is then constant along the path, and inference is a single deterministic update from the degraded latent.

\textbf{Latent-space flow matching.} Let $x_{\text{deg}} \in \mathbb{R}^{H \times W \times 3}$ be a decompressed image scaled to $[-1,1]$ at quality factor $\text{QF} \in [1, 30]$, and let $x_{\text{gt}}$ be its clean target. We encode both with the frozen variational autoencoder (VAE)~\citep{kingmaAutoEncodingVariationalBayes2022} of Stable Diffusion~3~\citep{esserScalingRectifiedFlow2024}. This gives $z_{\text{deg}} = \mathcal{E}(x_{\text{deg}})$ and $z_{\text{clean}} = \mathcal{E}(x_{\text{gt}})$. We define a linear path between the two latents:
\begin{equation}
z_t = (1 - t)\, z_{\text{deg}} + t \cdot z_{\text{clean}}, \quad t \in [0, 1],
\label{eq:flow_interpolation}
\end{equation}
with constant target velocity $v_{\text{gt}} = z_{\text{clean}} - z_{\text{deg}}$. Because the path starts at the degraded latent rather than at noise, a single evaluation at $t{=}0$ gives a direct estimate of the clean latent. The image prior therefore comes from the pretrained backbone instead of sampling, so QO-Flow acts as a deterministic one-step predictor.

\textbf{Compression-aware conditioning.} Artifacts vary with the quality factor. At low QF, blocking and ringing remove mid- and high-frequency content, while at moderate QF only fine textures are lost. We therefore condition QO-Flow on the QF, which can be read from the JPEG quantization table at inference with negligible cost (Appendix~\ref{app:qf_estimation}). A multilayer perceptron (MLP) with SiLU activations~\citep{elfwingSigmoidWeightedLinearUnits2017} maps a 64-band sinusoidal encoding~\citep{vaswaniAttentionAllYou2017} of $q=\text{QF}/30$ to a single embedding. This embedding is repeated over the 77 text-token positions of the joint attention to form $\mathbf{c}_{\text{QF}}$, while the pooled embedding $\mathbf{c}_{\text{QF}}^{\text{pooled}}$ modulates adaptive layer normalization~\citep{esserScalingRectifiedFlow2024, peeblesScalableDiffusionModels2023}.

\textbf{Efficient adaptation via LoRA.} Instead of fine-tuning the full 2.1B-parameter SD3-medium backbone, we apply LoRA~\citep{huLoRALowRankAdaptation2021} to the image-stream attention projections ($W_Q, W_K, W_V, W_O$) in the 24 joint attention blocks:
\begin{equation}
h = W_0 x + \tfrac{\alpha}{r}\, B(Ax), \quad r{=}64,\; \alpha{=}128.
\label{eq:lora}
\end{equation}
LoRA and the QF encoder together account for 1.4\% of the Stage~1 parameters. The rest of the backbone and the VAE are frozen, which preserves the pretrained prior.

\textbf{Training objective.} The main objective is the flow matching loss:
\begin{equation}
\mathcal{L}_{\text{FM}} = \mathbb{E}_{t, z_t}\bigl[\|v_\theta(z_t, t, \mathbf{c}_{\text{QF}}, \mathbf{c}_{\text{QF}}^{\text{pooled}}) - v_{\text{gt}}\|_2^2\bigr],
\label{eq:fm_loss}
\end{equation}
where $t \sim \mathcal{U}(\epsilon, 1{-}\epsilon)$ with $\epsilon{=}10^{-5}$. The VAE decoder can turn small latent errors into visible pixel artifacts~\citep{yiFineStructurePreserved2025}, so we add a perceptual term on the decoded output:
\begin{equation}
\mathcal{L}_{\text{Stage1}} = \mathcal{L}_{\text{FM}} + \lambda_{1}\, \mathbb{E}_{t, z_t}\bigl[\mathcal{P}\bigl(\mathcal{D}(z_{\text{deg}} + v_\theta), \, x_{\text{gt}}\bigr)\bigr],
\label{eq:stage1_loss}
\end{equation}
where $v_\theta$ is the prediction at the sampled $(z_t, t)$, $\mathcal{P}$ sums LPIPS~\citep{zhangUnreasonableEffectivenessDeep2018} and an $L_1$ distance between VGG features~\citep{simonyanVGG}, and $\lambda_{1}{=}0.2$.

\textbf{One-step inference.} At test time, we set $t{=}0$ and compute $z^{\text{pred}} = z_{\text{deg}} + v_\theta(z_{\text{deg}}, 0, \mathbf{c}_{\text{QF}}, \mathbf{c}_{\text{QF}}^{\text{pooled}})$ in one forward pass. Decoding this latent gives the structural prior $\hat{x}_{\text{FM}} = \mathcal{D}(z^{\text{pred}})$. Among the step counts we tested (1, 2, 5, and 10), one step achieves the lowest LPIPS and latency, while two steps raise PSNR by about 0.1~dB with higher latency (Table~\ref{tab:nfe}).

\subsection{Flow-Conditioned Detail Refinement (Stage~2)}

Stage~1 recovers global structure, but its output carries two kinds of error. The pretrained prior can synthesize plausible textures that differ from the ground truth, and the VAE decoder adds pixel-level error. Both errors lower pixel fidelity. Stage~2 reduces them through a deterministic pixel-space refinement that adds a correction to the decompressed input.

\textbf{Residual on the decompressed input.} FCDR receives two signals. The decompressed input $x_{\text{deg}}$ is faithful to the ground truth at low frequencies but lacks detail. The Stage~1 output $\hat{x}_{\text{FM}}$ is richer in structure but may deviate from the ground truth. FCDR predicts a residual that is added to the input:
\begin{equation}
\hat{x} = \text{clip}_{[-1,1]}\bigl(x_{\text{deg}} + 
\Delta x_\phi(x_{\text{deg}},\, \hat{x}_{\text{FM}})\bigr).
\label{eq:residual_formulation}
\end{equation}
This formulation expresses $\hat{x}$ as the observation plus a learned correction, so the network models only the residual over $x_{\text{deg}}$. In pipelines such as DiffBIR and SUPIR~\citep{linDiffBIRBlindImage2024a, yuScalingExcellencePracticing2024}, the generative model produces the final output. FCDR instead uses $\hat{x}_{\text{FM}}$ as guidance, and its output starts from the decompressed input.

\begin{table*}[t]
\centering
\caption{\textbf{Fidelity and perceptual similarity under severe compression.} JPEG denotes the decompressed input and is not ranked. Best results are in \textcolor{red}{\textbf{bold red}}, second-best are in \textcolor{blue}{\underline{underlined blue}}, and third-best are marked with $^{*}$.}
\setlength{\tabcolsep}{5pt}
\resizebox{\textwidth}{!}{%
\normalsize
\begin{tabular}{lcccccccccccc}
\toprule
 & \multicolumn{6}{c}{\textbf{LIVE1}} & \multicolumn{6}{c}{\textbf{DIV2K-Val}} \\
\cmidrule(lr){2-7} \cmidrule(lr){8-13}
 & \multicolumn{3}{c}{\textbf{QF=1}} & \multicolumn{3}{c}{\textbf{QF=5}} & \multicolumn{3}{c}{\textbf{QF=1}} & \multicolumn{3}{c}{\textbf{QF=5}} \\
\cmidrule(lr){2-4} \cmidrule(lr){5-7} \cmidrule(lr){8-10} \cmidrule(lr){11-13}
\textbf{Methods} & PSNR$\uparrow$ & SSIM$\uparrow$ & LPIPS$\downarrow$ & PSNR$\uparrow$ & SSIM$\uparrow$ & LPIPS$\downarrow$ & PSNR$\uparrow$ & SSIM$\uparrow$ & LPIPS$\downarrow$ & PSNR$\uparrow$ & SSIM$\uparrow$ & LPIPS$\downarrow$ \\
\midrule
JPEG & 20.89 & 0.534 & 0.552 & 23.10 & 0.640 & 0.438 & 21.78 & 0.583 & 0.548 & 24.17 & 0.680 & 0.447 \\
\midrule
FBCNN & \second{21.71} & \second{0.572} & 0.503 & \second{24.89} & \best{0.712} & 0.373\rlap{$^{*}$} & 22.69\rlap{$^{*}$} & \second{0.626} & 0.480\rlap{$^{*}$} & 26.24\rlap{$^{*}$} & \best{0.760} & 0.344\rlap{$^{*}$} \\
JDEC & 20.76 & 0.541 & 0.522 & 23.78 & 0.690 & 0.412 & 21.61 & 0.586 & 0.497 & 24.90 & 0.736 & 0.380 \\
PromptCIR & 21.24 & 0.550 & 0.532 & 24.85\rlap{$^{*}$} & \second{0.707} & 0.381 & 22.19 & 0.596 & 0.517 & \second{26.28} & \second{0.758} & 0.355 \\
SwinIR & 21.10 & 0.557\rlap{$^{*}$} & 0.534 & 24.17 & 0.692 & 0.414 & 22.04 & 0.606 & 0.516 & 25.33 & 0.738 & 0.394 \\
DiffBIR & 18.66 & 0.441 & 0.490\rlap{$^{*}$} & 19.36 & 0.471 & 0.400 & 19.79 & 0.472 & 0.483 & 20.39 & 0.504 & 0.417 \\
CODiff & 21.58\rlap{$^{*}$} & 0.555 & \best{0.316} & 22.81 & 0.612 & \best{0.206} & \second{22.84} & 0.624\rlap{$^{*}$} & \best{0.301} & 24.50 & 0.682 & \best{0.208} \\
\midrule
\textbf{FDIR (ours)} & \best{22.94} & \best{0.604} & \second{0.386} & \best{24.95} & 0.702\rlap{$^{*}$} & \second{0.279} & \best{24.11} & \best{0.668} & \second{0.349} & \best{26.41} & 0.754\rlap{$^{*}$} & \second{0.257} \\
\bottomrule
\end{tabular}%
}
\label{tab:extreme_compression}
\end{table*}

\begin{table}[t]
\centering
\caption{FID and DISTS on DIV2K-Val. Marks follow Table~\ref{tab:extreme_compression}.}
\setlength{\tabcolsep}{3pt}
\small
\begin{tabular}{lcccc}
\toprule
 & \multicolumn{2}{c}{\textbf{QF=1}} & \multicolumn{2}{c}{\textbf{QF=5}} \\
\cmidrule(lr){2-3} \cmidrule(lr){4-5}
\textbf{Methods} & FID$\downarrow$ & DISTS$\downarrow$ & FID$\downarrow$ & DISTS$\downarrow$ \\
\midrule
JPEG & 109.2 & 0.407 & 68.8 & 0.318 \\
\midrule
FBCNN & 94.9 & 0.325 & 44.7 & 0.208 \\
JDEC & 143.6 & 0.322 & 79.7 & 0.228 \\
PromptCIR & 101.9 & 0.372 & 45.0 & 0.207 \\
SwinIR & 108.5 & 0.352 & 68.1 & 0.222 \\
DiffBIR & 109.7 & 0.223 & 71.3 & 0.187 \\
CODiff & \best{64.5} & \best{0.149} & \best{33.0} & \best{0.101} \\
\midrule
\textbf{FDIR (ours)} & \second{68.8} & \second{0.217} & \second{37.9} & \second{0.163} \\
\bottomrule
\end{tabular}%
\label{tab:fid_comparison}
\end{table}

\begin{figure*}[t]
\centering
\includegraphics[width=\textwidth]{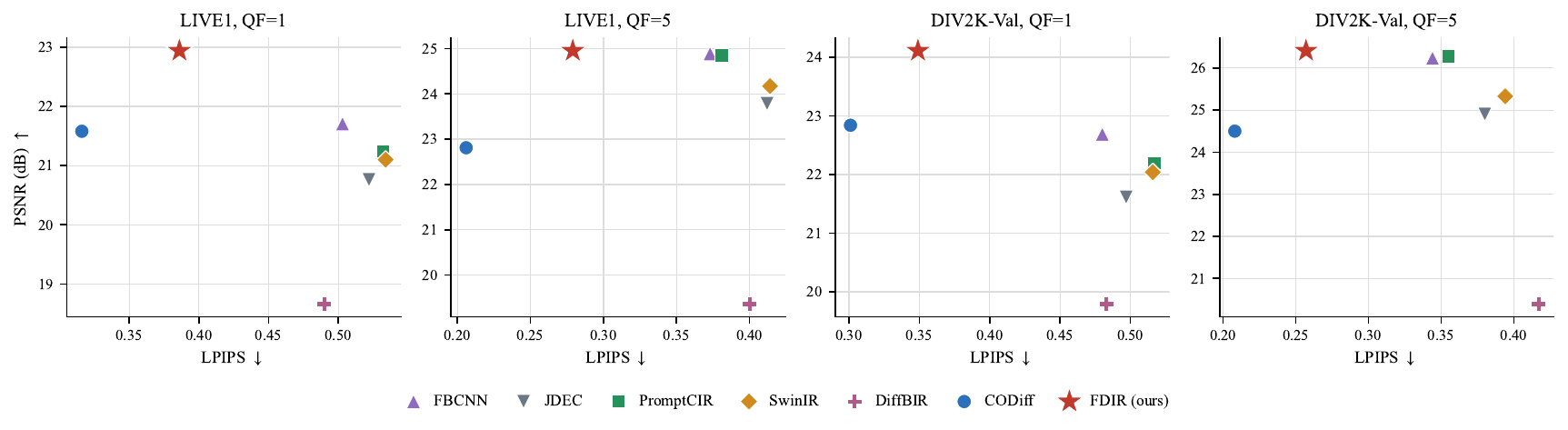}
\caption{\textbf{PSNR versus LPIPS under severe compression.} Each point is one method (Table~\ref{tab:extreme_compression}). The upper left is better.}
\label{fig:tradeoff}
\end{figure*}

\textbf{Architecture.} $\Delta x_\phi$ takes the concatenation $[x_{\text{deg}},\, \hat{x}_{\text{FM}}] \in \mathbb{R}^{H \times W \times 6}$. A fusion module with convolution, group normalization (GN)~\citep{wuGroupNormalization2018}, and SiLU layers first merges the two inputs. A U-Net~\citep{ronnebergerUNet} with GN and SiLU blocks then predicts the residual (44M parameters in total). Its receptive field exceeds 100\,px and covers many $8{\times}8$ JPEG blocks.

\textbf{Training objective.} FCDR is trained with three terms:
\begin{equation}
\mathcal{L}_{\text{FCDR}} = \mathcal{L}_{\text{rec}} + \lambda_{\text{perc}}\,
\mathcal{L}_{\text{perc}} + \mathcal{L}_{\text{struct}}.
\label{eq:fcdr_loss}
\end{equation}
The reconstruction term $\mathcal{L}_{\text{rec}} = \|\hat{x} - x_{\text{gt}}\|_1 + 0.5\,\|\hat{x} - x_{\text{gt}}\|_2^2$ targets fidelity. The perceptual term $\mathcal{L}_{\text{perc}} = \text{LPIPS}(\hat{x}, x_{\text{gt}})$ promotes perceptual similarity~\citep{zhangUnreasonableEffectivenessDeep2018}. The structure term is
\begin{equation}
\mathcal{L}_{\text{struct}} = \lambda_{e}\,\bigl\| |\nabla^2 \hat{x}| - |\nabla^2 x_{\text{gt}}| \bigr\|_1 + \lambda_{c}\,\|\hat{x} - \hat{x}_{\text{FM}}\|_1 ,
\label{eq:struct_loss}
\end{equation}
where the first part matches Laplacian edge maps of the luminance channel and the second keeps the output consistent with the structural prior. We set $\lambda_{\text{perc}}{=}0.01$. Over training, $\lambda_{e}$ increases linearly from 0.2 toward 0.3 and $\lambda_{c}$ decreases from 0.2 toward 0.1. The five coefficients $(1,\, 0.5,\, \lambda_{\text{perc}},\, \lambda_{e},\, \lambda_{c})$ are divided by their sum. With these weights, the two reconstruction terms make up about two-thirds of the loss value during training and the perceptual term about 2\%, reflecting the focus on fidelity. Appendix~\ref{app:stage2_weight} reports results with other values of $\lambda_{\text{perc}}$.

\begin{table*}[t]
\centering
\caption{\textbf{Machine preference at QF=20 and 30.} Each score is averaged over several pretrained models per task. JPEG denotes the decompressed input and is not ranked. Best in \textcolor{red}{\textbf{bold red}}, second-best in \textcolor{blue}{\underline{underlined blue}}.}
\scriptsize
\resizebox{\textwidth}{!}{%
\begin{tabular}{l|cc|cc|cc|cc|c|c}
\toprule
\multirow{2}{*}{\textbf{Methods}} & \multicolumn{4}{c|}{\textbf{Detection (COCO)}} & \multicolumn{4}{c|}{\textbf{Segmentation (CUB-200)}} & \multicolumn{2}{c}{\textbf{Retrieval (ImageNet)}} \\
\cmidrule(r){2-5} \cmidrule(lr){6-9} \cmidrule(l){10-11}
& \multicolumn{2}{c|}{\textbf{QF=20}} & \multicolumn{2}{c|}{\textbf{QF=30}} & \multicolumn{2}{c|}{\textbf{QF=20}} & \multicolumn{2}{c|}{\textbf{QF=30}} & \multicolumn{1}{c|}{\textbf{QF=20}} & \multicolumn{1}{c}{\textbf{QF=30}} \\
\cmidrule(lr){2-3} \cmidrule(lr){4-5} \cmidrule(lr){6-7} \cmidrule(lr){8-9} \cmidrule(lr){10-10} \cmidrule(l){11-11}
  & mAP$\uparrow$ & mAP$_{50}\uparrow$ & mAP$\uparrow$ & mAP$_{50}\uparrow$ & mIoU$\uparrow$ & Dice$\uparrow$ & mIoU$\uparrow$ & Dice$\uparrow$ & Top-1$\uparrow$ & Top-1$\uparrow$ \\
\midrule
JPEG & 44.87 & 66.32 & 50.30 & 72.03 & 79.56 & 87.08 & 78.85 & 86.42 & 68.39 & 70.17 \\
\midrule
FBCNN & 50.25 & 72.26 & 51.94 & 74.59 & 80.29 & 87.84 & 78.59 & 86.21 & 68.22 & 69.22 \\
JDEC & \best{52.40} & \best{73.54} & \best{53.82} & \best{75.31} & 79.91 & 87.47 & 78.76 & 86.38 & 68.00 & 69.78 \\
PromptCIR & 47.08 & 70.98 & 48.16 & 72.36 & 80.33 & 87.83 & 78.94 & 86.56 & 68.95 & 69.72 \\
SwinIR & 48.26 & 69.48 & 51.53 & 73.27 & 80.17 & 87.68 & 79.15 & 86.76 & 68.61 & 69.44 \\
DiffBIR & 42.32 & 60.74 & 41.80 & 60.51 & 80.23 & 87.80 & 78.46 & 86.53 & 61.78 & 61.72 \\
CODiff & 51.49 & 73.17 & 51.01 & 73.07 & \best{81.68} & \best{88.93} & \best{80.60} & \best{87.96} & \best{71.05} & \best{72.28} \\
\midrule
\textbf{FDIR (ours)} & \second{52.17} & \second{73.38} & \second{53.36} & \second{74.89} & \second{80.47} & \second{87.94} & \second{79.42} & \second{86.96} & \second{69.05} & \second{70.67} \\
\bottomrule
\end{tabular}%
}
\label{tab:downstream_tasks}
\end{table*}

\begin{figure*}[t]
\centering
\includegraphics[width=\textwidth]{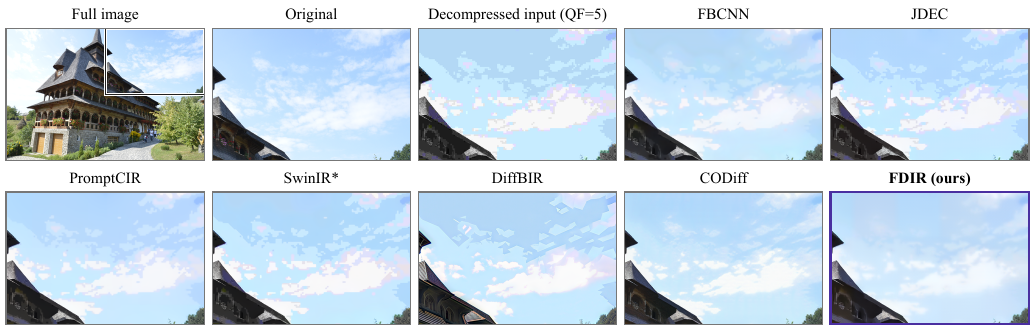}
\caption{\textbf{Visual comparison at QF=5} on DIV2K-Val. SwinIR$^*$ uses its QF=10 model. DiffBIR uses tiled inference.}
\label{fig:visual_comparison}
\end{figure*}

\section{Experiments}
\label{sec:experiments}

\textbf{Evaluation criteria.} We assess restoration on three criteria. Fidelity measures pixel agreement with the ground truth through PSNR and SSIM~\citep{wangImageQualityAssessment2004}. It matters most where invented details are harmful, such as false textures that change the image content~\citep{cohenLooksTooGood2024}. Human perception combines perceptual similarity to the reference, measured by LPIPS~\citep{zhangUnreasonableEffectivenessDeep2018} and DISTS~\citep{dingImageQualityAssessment2020}, with distribution-level realism, measured by FID~\citep{heuselFID}. Machine preference measures how well detection, segmentation, and retrieval models perform on the restored images, following the task types of \citet{liImageQualityAssessment2025}.

\textbf{Setup.} We use two protocols. Fidelity and perception are evaluated under \emph{severe compression} (QF=1 and 5), where quantization removes most mid- and high-frequency content. Machine preference is evaluated at QF=20 and 30, which are closer to practical use. Our DF2K dataset contains 3,450 images (800 from DIV2K~\citep{agustssonNTIRE2017Challenge2017} and 2,650 from Flickr2K~\citep{timofteNTIRE2017Methods}), split into 3,105 for training and 345 for validation. During training, we sample $256 \times 256$ patches with $\text{QF} \sim \mathcal{U}(1, 30)$. Evaluation uses LIVE1~\citep{sheikhLiveImageQuality2005a} (29 images) and DIV2K-Val~\citep{agustssonNTIRE2017Challenge2017} (100 images). We compare with six baselines: the CNN method FBCNN~\citep{jiangFlexibleBlindJPEG2021}, the DCT-domain decoder JDEC~\citep{hanJDECJPEGDecoding2024}, the Transformer methods SwinIR~\citep{liangSwinIRImageRestoration2021} and PromptCIR~\citep{liPromptCIRBlindCompressed2024}, and the generative methods DiffBIR~\citep{linDiffBIRBlindImage2024a} and CODiff~\citep{guoCompressionAwareOneStepDiffusion2025}.

\textbf{Implementation details.} QO-Flow fine-tunes SD3-medium with LoRA ($r{=}64$, $\alpha{=}128$) for 125K steps using AdamW~\citep{loshchilovAdamW} (learning rate 5$\times$10$^{-5}$, weight decay 10$^{-6}$) with an effective batch size of 20. We use the checkpoint with the highest validation PSNR. FCDR is trained from scratch for 50 epochs with AdamW (learning rate 2$\times$10$^{-4}$, weight decay 10$^{-5}$) and an effective batch size of 24. Stage~1 is trained first, and Stage~2 trains on its frozen outputs. Large images are processed in overlapping $256 \times 256$ tiles.

\subsection{Comparison with Existing Methods}

\textbf{Fidelity.} Table~\ref{tab:extreme_compression} reports results under severe compression. FDIR achieves the highest PSNR in all four settings. At QF=1, it leads the second-best method by 1.23~dB on LIVE1 and 1.27~dB on DIV2K-Val, and also obtains the highest SSIM. At QF=5, its SSIM is within 0.010 of the best regression method. Compared with CODiff, the strongest perceptual baseline, FDIR is 1.3 to 2.1~dB higher in PSNR. FDIR thus has higher PSNR than every regression method, while its Stage~1 prior improves perceptual quality on top of this fidelity (Sec.~\ref{sec:ablation}).

\textbf{Perceptual similarity.} FDIR places second on LPIPS in all four settings, behind only CODiff. Its LPIPS is 0.087 to 0.131 lower than the best result among the other five methods. In every setting, it is closer to CODiff than to any other method. The regression methods produce smoother outputs, reflected in their much higher LPIPS.

\textbf{FID and DISTS.} Table~\ref{tab:fid_comparison} reports FID and DISTS on DIV2K-Val. FDIR is second to CODiff on both metrics at both quality factors. Its FID is 4.3 and 4.9 higher than that of CODiff at QF=1 and 5, whereas every other method is at least 30.4 and 11.7 higher. Figure~\ref{fig:tradeoff} plots PSNR against LPIPS. In every panel, FDIR lies at the top, between CODiff and the other five methods along the LPIPS axis.

\textbf{Visual comparison.} In Figure~\ref{fig:visual_comparison}, the regression methods leave visible blocking and banding in the sky, and DiffBIR synthesizes structures absent from the ground truth. FDIR removes these artifacts while preserving the original colors and shapes.

\begin{table*}[t]
\centering
\caption{\textbf{Ablation studies.} (Left) Fidelity and perceptual similarity (P: PSNR, S: SSIM, L: LPIPS) on DIV2K-Val. (Right) Machine preference at QF=10. All trained variants use the same training setting as FDIR.}
\label{tab:ablation}
\vskip 0.05in
\begin{minipage}[t]{0.57\textwidth}
\centering
\resizebox{\linewidth}{!}{%
\small
\begin{tabular}{l|ccc|ccc|ccc}
\toprule
& \multicolumn{3}{c|}{\textbf{QF=1}} & \multicolumn{3}{c|}{\textbf{QF=5}} & \multicolumn{3}{c}{\textbf{QF=10}} \\
\cmidrule(lr){2-4} \cmidrule(lr){5-7} \cmidrule(l){8-10}
\textbf{Config.} & P$\uparrow$ & S$\uparrow$ & L$\downarrow$ & P$\uparrow$ & S$\uparrow$ & L$\downarrow$ & P$\uparrow$ & S$\uparrow$ & L$\downarrow$ \\
\midrule
JPEG & 21.78 & 0.583 & 0.548 & 24.17 & 0.680 & 0.447 & 27.10 & 0.781 & 0.323 \\
\midrule
w/o Stage 1 & 24.15 & 0.668 & 0.372 & 26.46 & 0.756 & 0.272 & 29.22 & 0.838 & 0.185 \\
w/o Stage 2 & 23.31 & 0.641 & 0.350 & 25.37 & 0.720 & 0.251 & 27.67 & 0.797 & 0.167 \\
\midrule
w/o $\mathcal{L}_{\text{rec}}$ & 23.26 & 0.637 & 0.334 & 25.35 & 0.718 & 0.239 & 27.71 & 0.797 & 0.160 \\
w/o $\mathcal{L}_{\text{perc}}$ & 24.18 & 0.673 & 0.417 & 26.49 & 0.759 & 0.321 & 29.27 & 0.840 & 0.233 \\
w/o $\mathcal{L}_{\text{struct}}$ & 24.21 & 0.673 & 0.353 & 26.53 & 0.760 & 0.258 & 29.31 & 0.840 & 0.176 \\
\midrule
\textbf{FDIR} & 24.11 & 0.668 & 0.349 & 26.41 & 0.754 & 0.257 & 29.19 & 0.836 & 0.179 \\
\bottomrule
\end{tabular}%
}
\end{minipage}%
\hfill
\begin{minipage}[t]{0.41\textwidth}
\centering
\resizebox{\linewidth}{!}{%
\small
\begin{tabular}{l|cc|cc|c}
\toprule
& \multicolumn{2}{c|}{\textbf{Det.}} & \multicolumn{2}{c|}{\textbf{Seg.}} & \textbf{Ret.} \\
\cmidrule(lr){2-3} \cmidrule(lr){4-5} \cmidrule(l){6-6}
\textbf{Config.} & mAP & mAP$_{50}$ & mIoU & Dice & Top-1 \\
\midrule
JPEG & 25.64 & 43.93 & 70.70 & 78.95 & 64.05 \\
\midrule
w/o Stage 1 & 43.60 & 64.81 & 78.01 & 86.11 & 67.00 \\
w/o Stage 2 & 43.49 & 64.74 & 78.87 & 86.93 & 66.89 \\
\midrule
w/o $\mathcal{L}_{\text{rec}}$ & 43.73 & 64.88 & 78.95 & 86.96 & 67.28 \\
w/o $\mathcal{L}_{\text{perc}}$ & 44.44 & 65.32 & 78.36 & 86.45 & 67.17 \\
w/o $\mathcal{L}_{\text{struct}}$ & 44.50 & 65.17 & 78.33 & 86.41 & 66.61 \\
\midrule
\textbf{FDIR} & 44.81 & 65.99 & 78.06 & 86.12 & 67.33 \\
\bottomrule
\end{tabular}%
}
\end{minipage}%
\end{table*}

\subsection{Machine Preference Evaluation}

\textbf{Setting.} We evaluate each task across multiple pretrained models to reduce model bias. (1)~\emph{Object detection}: Faster R-CNN~\citep{renFasterRCNNRealTime2016}, FCOS~\citep{tianFCOSFullyConvolutional2019}, and RetinaNet~\citep{linFocalLoss} on 500 COCO validation images~\citep{linMicrosoftCOCOCommon2015}, reporting mean average precision (mAP) and mAP$_{50}$. (2)~\emph{Segmentation}: DeepLabV3-ResNet101~\citep{chenDeepLabv3, heDeepResidualLearning2015}, Mask R-CNN-ResNet50~\citep{heMaskRCNN2018}, and YOLOv8m~\citep{jocherYOLOv8} on 300 CUB-200-2011 images~\citep{wahCaltechUCSDBirds2002011Dataset}, reporting mean intersection over union (mIoU) and Dice. (3)~\emph{Image retrieval}: embeddings from CLIP-ViT-B/16~\citep{radfordCLIP}, DINOv2-ViT-B/14~\citep{oquabDINOv2LearningRobust2024}, EfficientNet-B4~\citep{tanEfficientNetRethinkingModel2020}, ResNet-50~\citep{heDeepResidualLearning2015}, Swin-B~\citep{liuSwinTransformerHierarchical2021}, and ViT-B/16~\citep{dosovitskiyImageWorth16x162021} on 300 ImageNet queries~\citep{dengImageNetLargescaleHierarchical2009} against a 20K gallery, reporting Top-1 accuracy.

\textbf{Results.} Table~\ref{tab:downstream_tasks} shows that FDIR ranks second on all ten downstream metrics and obtains the best average rank (2.0, against 2.2 for CODiff). Other methods perform strongly on only a subset of the tasks. JDEC leads detection but ranks fifth to seventh on segmentation. CODiff leads segmentation and retrieval but ranks third to fifth on detection, where FDIR is 0.21 to 2.35 points higher. FDIR is the only method in the top two on every metric, and it is higher than FBCNN, PromptCIR, SwinIR, and DiffBIR on all six segmentation and retrieval metrics. It also improves over the decompressed input on all ten metrics, for example by 7.30 mAP for detection at QF=20. These results indicate that a fidelity-anchored model remains useful for all three tasks.

\begin{figure*}[t]
\centering
\includegraphics[width=\textwidth]{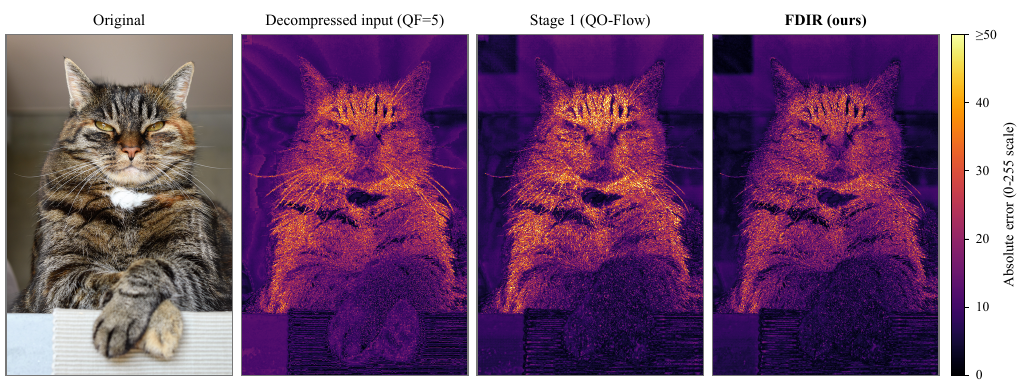}
\caption{\textbf{Error maps.} Left to right: original, and absolute-error maps of the decompressed input, the Stage~1 output, and FDIR on DIV2K-Val at QF=5. Errors are on the 0--255 scale and clipped at 50.}
\label{fig:stage_progression}
\end{figure*}

\subsection{Ablation Studies}
\label{sec:ablation}
\textbf{Role of the two stages.} Table~\ref{tab:ablation} isolates each stage. The Stage~1 output alone (w/o Stage~2) is 0.80, 1.04, and 1.52~dB below full FDIR in PSNR at QF=1, 5, and 10. Adding Stage~2 recovers this fidelity and raises SSIM by 0.027 to 0.039, while keeping LPIPS within 0.012 of the Stage~1 output. The error maps in Figure~\ref{fig:stage_progression} show this on one image: errors in the Stage~1 output are concentrated in textured regions, and FDIR reduces them. Removing Stage~1 trains FCDR directly on the decompressed input, without the prior or its consistency term. In that setting, PSNR remains within 0.05~dB of full FDIR. However, LPIPS rises by 0.023, 0.015, and 0.006 at QF=1, 5, and 10, and detection drops by 1.21 mAP at QF=10. The prior thus contributes mainly to perceptual similarity, especially at the lowest quality factors, and to detection accuracy. The variant without $\mathcal{L}_{\text{struct}}$, which keeps the prior as input but drops the edge and consistency terms, retains this LPIPS gain.

\textbf{Loss terms.} Without $\mathcal{L}_{\text{rec}}$, PSNR and SSIM fall to the level of the Stage~1 output, and LPIPS improves by 0.015 to 0.019. $\mathcal{L}_{\text{rec}}$ therefore shifts FCDR from the prior toward the ground truth. Without $\mathcal{L}_{\text{perc}}$, LPIPS increases by 0.054 to 0.068 while PSNR rises by 0.07 to 0.08~dB. These two terms act on opposite sides of the fidelity-perception trade-off, and the full model sits between them. Removing both parts of $\mathcal{L}_{\text{struct}}$ raises PSNR by 0.10 to 0.12~dB and changes LPIPS by at most 0.004, but lowers detection mAP by 0.31 and retrieval Top-1 by 0.72 at QF=10.

\textbf{Trade-off among the three criteria.} Theory states that distortion and the divergence from the natural-image distribution cannot both be minimized~\citep{blauPerceptionDistortionTradeoff2018}, and this result has been extended to classification error~\citep{liuClassificationDistortionPerception2019}. Our comparison shows the same pattern. No compared method leads on all three criteria. CODiff, the method with the lowest FID, has lower PSNR than FDIR in all four settings. FBCNN and PromptCIR, the two methods closest to FDIR in PSNR at QF=5, have FID at least 6.8 higher. In this comparison, FDIR is first on PSNR in all four settings and second on every perceptual and downstream metric. Figure~\ref{fig:radar} summarizes the mean rank of each method per criterion. The mean ranks of FDIR are 1.5 for fidelity, 2.0 for perception, and 2.0 for machine preference. Every other method has a mean rank of 4.5 or worse on at least one criterion. Retraining FCDR with perceptual weights up to 0.3 lowers LPIPS and PSNR together (Appendix~\ref{app:stage2_weight}), so the trade-off remains. FDIR operates on the fidelity-first side of this trade-off.

\begin{table*}[t]
\begin{minipage}[t]{0.47\textwidth}
\centering
\caption{Inference latency on $256 \times 256$ inputs.}
\label{tab:latency_comparison}
\small
\begin{tabular}{lcc}
\toprule
\textbf{Method} & ms$\downarrow$ & FPS$\uparrow$ \\
\midrule
FBCNN & 9.3 & 107.4 \\
JDEC & 43.5 & 23.0 \\
PromptCIR & 282.3 & 3.5 \\
SwinIR & 233.3 & 4.3 \\
DiffBIR & 2361.5 & 0.4 \\
CODiff & 120.2 & 8.3 \\
\midrule
\textbf{FDIR (ours)} & 49.8 & 20.1 \\
\bottomrule
\end{tabular}%
\end{minipage}\hfill
\begin{minipage}[t]{0.47\textwidth}
\centering
\caption{\textbf{Number of inference steps} for the Stage~1 output (100 DIV2K-Val $256\times256$ center crops). P: PSNR, L: LPIPS.}
\label{tab:nfe}
\small
\begin{tabular}{c|cc|cc|c}
\toprule
& \multicolumn{2}{c|}{\textbf{QF=1}} & \multicolumn{2}{c|}{\textbf{QF=5}} & \\
\textbf{Steps} & P$\uparrow$ & L$\downarrow$ & P$\uparrow$ & L$\downarrow$ & ms$\downarrow$ \\
\midrule
\textbf{1 (ours)} & 22.34 & 0.345 & 24.35 & 0.238 & 45.5 \\
2 & 22.44 & 0.364 & 24.47 & 0.255 & 72.7 \\
5 & 22.32 & 0.382 & 24.35 & 0.280 & 154.4 \\
10 & 22.15 & 0.395 & 24.16 & 0.294 & 290.1 \\
\bottomrule
\end{tabular}
\end{minipage}
\end{table*}

\subsection{Efficiency}

Table~\ref{tab:latency_comparison} reports end-to-end latency on $256\times256$ inputs. FDIR takes 49.8~ms per image on an RTX 4090 in half precision (fp16). This is 2.4$\times$ faster than one-step CODiff, 47$\times$ faster than multi-step DiffBIR, and faster than the Transformer methods SwinIR and PromptCIR. Of its 2.19B parameters, only 73.5M are trained, and the pretrained backbone and VAE are frozen. FDIR is also trained on only 3.1K DF2K images, fewer than CODiff (88K)~\citep{guoCompressionAwareOneStepDiffusion2025} and DiffBIR (15M)~\citep{linDiffBIRBlindImage2024a}.

\raggedbottom
\section{Conclusion}
\label{sec:conclusion}
Restored images serve both human viewers and vision models, so we evaluate them on pixel fidelity, human perception, and machine preference. We presented FDIR, a fidelity-anchored framework for decompressed image restoration. A one-step flow model generates a structural prior, and a refinement network uses it to predict a residual on the decompressed input. The flow model requires a single forward pass, and only 73.5M parameters are trained. Under severe JPEG compression, FDIR attains the highest PSNR among the compared methods and ranks second only to CODiff on perceptual metrics. On detection, segmentation, and retrieval, it ranks second on every metric and has the best average rank. Fidelity and perceptual quality are known to conflict~\citep{blauPerceptionDistortionTradeoff2018, liuClassificationDistortionPerception2019}, and machine preference follows neither reliably~\citep{liImageQualityAssessment2025}. FDIR operates within this trade-off. It keeps fidelity first and remains in the top two on the other two criteria. A restoration model can therefore put fidelity first and still serve both human viewers and vision models.

\textbf{Limitations.} FDIR is designed for lossy compression. The main experiments use JPEG, and Appendix~\ref{app:webp} reports a model retrained on WebP. Its perceptual scores remain behind those of CODiff because it puts fidelity first. Directions for future work include extending the framework to other codecs and degradations, and letting users set the balance between fidelity and perception at inference time rather than by retraining.

\section*{Use of AI Tools}
We used Claude Code (Anthropic) and Codex (OpenAI) to assist with writing code and scripts, and Gemini (Google) and Claude (Anthropic) to polish the manuscript's language. The authors double-checked the reported numbers and figures and approved the final manuscript.

{
    \small
    \bibliographystyle{ieeenat_fullname}
    \bibliography{ref}
}
\clearpage
\appendix
\section{Implementation Details}
\label{app:implementation}

\textbf{Data.} Training uses 3,105 DF2K images and validation uses the remaining 345. Each training sample is a random $256{\times}256$ crop of the ground truth. It is encoded with OpenCV JPEG at a quality factor drawn uniformly from $[1, 30]$ and decoded back to RGB. LIVE1 and DIV2K-Val are encoded in the same way at the evaluated QF.

\textbf{Stage~1 (QO-Flow).} The backbone is SD3-medium. LoRA ($r{=}64$, $\alpha{=}128$) is applied to $W_Q, W_K, W_V, W_O$ of the image stream in all 24 joint attention blocks. The QF encoder maps $q=\text{QF}/30$ to a 64-band sinusoidal encoding and then to the conditioning embedding with an MLP. LoRA and the QF encoder have 29.5M trainable parameters. Training runs for 125K steps with AdamW (learning rate $5{\times}10^{-5}$, weight decay $10^{-6}$) and an effective batch size of 20. The perceptual term of Eq.~(\ref*{eq:stage1_loss}) uses $\lambda_1{=}0.2$. We keep the checkpoint with the highest validation PSNR.

\textbf{Stage~2 (FCDR).} FCDR has 44.0M parameters and is trained from scratch for 50 epochs with AdamW (learning rate $2{\times}10^{-4}$, weight decay $10^{-5}$) and an effective batch size of 24. Stage~1 is frozen during this stage. The edge weight $\lambda_e$ and the consistency weight $\lambda_c$ follow a linear schedule over the epochs. At the last epoch they reach 0.298 and 0.102. The edge term uses the luminance $Y = 0.299R + 0.587G + 0.114B$ and a four-neighbor Laplacian.

\textbf{Inference.} Images larger than $256{\times}256$ are split into $256{\times}256$ tiles with a 32-pixel overlap. Each tile passes through Stage~1 and Stage~2, and the overlaps are blended with linear weights. At inference, the QF can be read from the JPEG file (Sec.~\ref{app:qf_estimation}).

\section{Evaluation Metrics}
\label{app:quality_metrics}

All metrics are computed on RGB images. Restored images are rounded to 8 bits before scoring.

\textbf{PSNR.} PSNR measures pixel error between the restored image $\hat{x}$ and the ground truth $x_{\text{gt}}$:
\begin{equation}
    \text{PSNR}(\hat{x}, x_{\text{gt}}) = 10 \log_{10} \frac{255^2}{\text{MSE}(\hat{x}, x_{\text{gt}})},
\end{equation}
where the mean squared error is taken over height, width, and channels.

\textbf{SSIM~\citep{wangImageQualityAssessment2004}.} SSIM compares local luminance, contrast, and structure:
\begin{equation}
    \text{SSIM}(x, y) = \frac{(2\mu_x\mu_y + C_1)(2\sigma_{xy} + C_2)}{(\mu_x^2 + \mu_y^2 + C_1)(\sigma_x^2 + \sigma_y^2 + C_2)},
\end{equation}
where $\mu$ and $\sigma^2$ are local means and variances, $\sigma_{xy}$ is the covariance, and $C_1, C_2$ are small constants. We compute it on the three channels with a data range of 255.

\textbf{LPIPS~\citep{zhangUnreasonableEffectivenessDeep2018}.} LPIPS is a weighted distance between deep features:
\begin{equation}
    \text{LPIPS}(\hat{x}, x_{\text{gt}}) = \sum_{\ell} \frac{1}{H_\ell W_\ell} \sum_{h,w} \| \mathbf{w}_\ell \odot (\phi_\ell(\hat{x}) - \phi_\ell(x_{\text{gt}})) \|_2^2,
\end{equation}
where $\phi_\ell$ is the normalized feature map at layer $\ell$ and $\mathbf{w}_\ell$ are learned channel weights. We use the AlexNet version.

\textbf{DISTS~\citep{dingImageQualityAssessment2020}.} DISTS combines structure and texture terms over VGG layers:
\begin{equation}
    \text{DISTS}(\hat{x}, x_{\text{gt}}) = \sum_{\ell} \bigl(\alpha_\ell \, d_{\text{struct}}^{(\ell)} + \beta_\ell \, d_{\text{texture}}^{(\ell)}\bigr),
\end{equation}
with learned weights $\alpha_\ell$ and $\beta_\ell$. It tolerates texture resampling and remains sensitive to structural change.

\textbf{FID~\citep{heuselFID}.} FID compares Gaussian fits of Inception-V3 pool features (2048 dimensions) for the restored and ground-truth sets:
\begin{equation}
    \text{FID} = \|\mu_{\hat{x}} - \mu_{x_{\text{gt}}}\|_2^2 + \text{Tr}\bigl(\Sigma_{\hat{x}} + \Sigma_{x_{\text{gt}}} - 2(\Sigma_{\hat{x}} \Sigma_{x_{\text{gt}}})^{1/2}\bigr).
\end{equation}
We report it on the 100 full-resolution DIV2K-Val images, following prior compression restoration work~\citep{guoCompressionAwareOneStepDiffusion2025}.

\section{Perceptual Weight of Stage~1}
\label{app:stage1_weight}

Table~\ref{tab:stage1_weight} evaluates the Stage~1 output alone for three values of $\lambda_1$ in Eq.~(\ref*{eq:stage1_loss}). The variant without the perceptual term has 0.072 higher LPIPS and 0.23~dB lower PSNR at QF=1. Raising $\lambda_1$ from 0.2 to 0.5 lowers LPIPS by 0.020 at QF=1, and PSNR changes by less than 0.1~dB. We use $\lambda_1{=}0.2$, which has the highest PSNR at QF=1 and 5.

\begin{table*}[t]
\centering
\caption{Stage~1 output on DIV2K-Val for different perceptual weights $\lambda_1$. The run without the perceptual term was trained separately with a different micro-batch size and checkpoint selection.}
\label{tab:stage1_weight}
\small
\begin{tabular}{lccccccccc}
\toprule
 & \multicolumn{3}{c}{\textbf{QF=1}} & \multicolumn{3}{c}{\textbf{QF=5}} & \multicolumn{3}{c}{\textbf{QF=10}} \\
\cmidrule(lr){2-4} \cmidrule(lr){5-7} \cmidrule(lr){8-10}
 & PSNR$\uparrow$ & SSIM$\uparrow$ & LPIPS$\downarrow$ & PSNR$\uparrow$ & SSIM$\uparrow$ & LPIPS$\downarrow$ & PSNR$\uparrow$ & SSIM$\uparrow$ & LPIPS$\downarrow$ \\
\midrule
w/o perceptual term & 23.08 & 0.619 & 0.422 & 25.11 & 0.707 & 0.297 & 27.37 & 0.791 & 0.195 \\
$\lambda_1{=}0.2$ (ours) & 23.31 & 0.641 & 0.350 & 25.37 & 0.720 & 0.251 & 27.67 & 0.797 & 0.167 \\
$\lambda_1{=}0.5$ & 23.23 & 0.637 & 0.330 & 25.35 & 0.721 & 0.240 & 27.73 & 0.803 & 0.163 \\
\bottomrule
\end{tabular}
\end{table*}

\section{Perceptual Weight of Stage~2}
\label{app:stage2_weight}

We retrain FCDR with $\lambda_{\text{perc}} \in \{0.01, 0.03, 0.1, 0.3\}$ in Eq.~(\ref*{eq:fcdr_loss}). Stage~1 is the same, and the other loss weights keep their ratios. Tables~\ref{tab:stage2_weight_div2k} and~\ref{tab:stage2_weight_live1} report the results. As $\lambda_{\text{perc}}$ increases, LPIPS and DISTS decrease and PSNR and SSIM decrease with them. On DIV2K-Val at QF=1, going from 0.01 to 0.3 lowers LPIPS from 0.349 to 0.280 and PSNR from 24.11 to 23.87~dB. FID varies by at most 2.1 across the four weights. The weight therefore moves FDIR along the fidelity--perception trade-off rather than improving both fidelity and perception. We use 0.01 to keep the focus on fidelity.

\begin{table*}[t]
\centering
\caption{FCDR retrained with different perceptual weights $\lambda_{\text{perc}}$ on DIV2K-Val. The first row is FDIR.}
\label{tab:stage2_weight_div2k}
\small
\begin{tabular}{lcccccccccc}
\toprule
 & \multicolumn{5}{c}{\textbf{QF=1}} & \multicolumn{5}{c}{\textbf{QF=5}} \\
\cmidrule(lr){2-6} \cmidrule(lr){7-11}
$\lambda_{\text{perc}}$ & PSNR$\uparrow$ & SSIM$\uparrow$ & LPIPS$\downarrow$ & DISTS$\downarrow$ & FID$\downarrow$ & PSNR$\uparrow$ & SSIM$\uparrow$ & LPIPS$\downarrow$ & DISTS$\downarrow$ & FID$\downarrow$ \\
\midrule
0.01 (ours) & 24.11 & 0.668 & 0.349 & 0.217 & 68.8 & 26.41 & 0.754 & 0.257 & 0.163 & 37.9 \\
0.03 & 24.08 & 0.665 & 0.319 & 0.202 & 68.9 & 26.37 & 0.752 & 0.228 & 0.151 & 37.5 \\
0.1 & 23.96 & 0.658 & 0.298 & 0.192 & 68.4 & 26.24 & 0.746 & 0.211 & 0.145 & 37.2 \\
0.3 & 23.87 & 0.652 & 0.280 & 0.183 & 66.8 & 26.16 & 0.742 & 0.198 & 0.139 & 36.3 \\
\bottomrule
\end{tabular}
\end{table*}

\begin{table}[t]
\centering
\caption{FCDR retrained with different perceptual weights $\lambda_{\text{perc}}$ on LIVE1.}
\label{tab:stage2_weight_live1}
\small
\setlength{\tabcolsep}{4pt}
\resizebox{\columnwidth}{!}{%
\begin{tabular}{lcccccc}
\toprule
 & \multicolumn{3}{c}{\textbf{QF=1}} & \multicolumn{3}{c}{\textbf{QF=5}} \\
\cmidrule(lr){2-4} \cmidrule(lr){5-7}
$\lambda_{\text{perc}}$ & PSNR$\uparrow$ & SSIM$\uparrow$ & LPIPS$\downarrow$ & PSNR$\uparrow$ & SSIM$\uparrow$ & LPIPS$\downarrow$ \\
\midrule
0.01 (ours) & 22.94 & 0.604 & 0.386 & 24.95 & 0.702 & 0.279 \\
0.03 & 22.91 & 0.601 & 0.354 & 24.91 & 0.699 & 0.246 \\
0.1 & 22.79 & 0.593 & 0.324 & 24.78 & 0.693 & 0.223 \\
0.3 & 22.70 & 0.587 & 0.304 & 24.69 & 0.689 & 0.209 \\
\bottomrule
\end{tabular}%
}
\end{table}

\section{Residual Anchor of Stage~2}
\label{app:residual_anchor}

Eq.~(\ref*{eq:residual_formulation}) adds the predicted residual to the decompressed input $x_{\text{deg}}$. An alternative adds it to the Stage~1 output $\hat{x}_{\text{FM}}$. We retrain FCDR with this alternative and keep all other settings unchanged (Table~\ref{tab:residual_anchor}). The two anchors give similar results. PSNR differs by at most 0.03~dB and LPIPS by at most 0.002. Anchoring to $x_{\text{deg}}$ gives slightly higher PSNR in all three settings, and we use it.

\begin{table*}[t]
\centering
\caption{Residual anchor of FCDR on DIV2K-Val. JPEG denotes the decompressed input.}
\label{tab:residual_anchor}
\small
\begin{tabular}{lccccccccc}
\toprule
 & \multicolumn{3}{c}{\textbf{QF=1}} & \multicolumn{3}{c}{\textbf{QF=5}} & \multicolumn{3}{c}{\textbf{QF=10}} \\
\cmidrule(lr){2-4} \cmidrule(lr){5-7} \cmidrule(lr){8-10}
Anchor & PSNR$\uparrow$ & SSIM$\uparrow$ & LPIPS$\downarrow$ & PSNR$\uparrow$ & SSIM$\uparrow$ & LPIPS$\downarrow$ & PSNR$\uparrow$ & SSIM$\uparrow$ & LPIPS$\downarrow$ \\
\midrule
JPEG & 21.78 & 0.583 & 0.548 & 24.17 & 0.680 & 0.447 & 27.10 & 0.781 & 0.323 \\
\midrule
$\hat{x}_{\text{FM}}$ & 24.09 & 0.667 & 0.347 & 26.38 & 0.754 & 0.257 & 29.16 & 0.836 & 0.179 \\
$x_{\text{deg}}$ (ours) & 24.11 & 0.668 & 0.349 & 26.41 & 0.754 & 0.257 & 29.19 & 0.836 & 0.179 \\
\bottomrule
\end{tabular}
\end{table*}

\begin{table*}[t]
\centering
\caption{WebP restoration on DIV2K-Val. WebP denotes the decompressed input.}
\label{tab:webp}
\small
\setlength{\tabcolsep}{3.2pt}
\begin{tabular}{lcccccccccccc}
\toprule
 & \multicolumn{4}{c}{\textbf{QF=1}} & \multicolumn{4}{c}{\textbf{QF=5}} & \multicolumn{4}{c}{\textbf{QF=10}} \\
\cmidrule(lr){2-5} \cmidrule(lr){6-9} \cmidrule(lr){10-13}
Method & PSNR$\uparrow$ & SSIM$\uparrow$ & LPIPS$\downarrow$ & DISTS$\downarrow$ & PSNR$\uparrow$ & SSIM$\uparrow$ & LPIPS$\downarrow$ & DISTS$\downarrow$ & PSNR$\uparrow$ & SSIM$\uparrow$ & LPIPS$\downarrow$ & DISTS$\downarrow$ \\
\midrule
WebP & 27.50 & 0.774 & 0.322 & 0.194 & 28.65 & 0.813 & 0.268 & 0.170 & 29.45 & 0.838 & 0.230 & 0.153 \\
\midrule
QO-Flow (Stage~1) & 27.33 & 0.768 & 0.198 & 0.147 & 28.24 & 0.802 & 0.160 & 0.121 & 28.81 & 0.823 & 0.138 & 0.105 \\
FDIR & 28.80 & 0.809 & 0.202 & 0.142 & 29.98 & 0.844 & 0.164 & 0.124 & 30.79 & 0.865 & 0.141 & 0.113 \\
\bottomrule
\end{tabular}
\end{table*}

\begin{table*}[t]
\centering
\caption{FDIR with two Stage~1 backbones on DIV2K-Val. JPEG denotes the decompressed input.}
\label{tab:backbone}
\small
\begin{tabular}{lccccccccc}
\toprule
 & \multicolumn{3}{c}{\textbf{QF=1}} & \multicolumn{3}{c}{\textbf{QF=5}} & \multicolumn{3}{c}{\textbf{QF=10}} \\
\cmidrule(lr){2-4} \cmidrule(lr){5-7} \cmidrule(lr){8-10}
Backbone & PSNR$\uparrow$ & SSIM$\uparrow$ & LPIPS$\downarrow$ & PSNR$\uparrow$ & SSIM$\uparrow$ & LPIPS$\downarrow$ & PSNR$\uparrow$ & SSIM$\uparrow$ & LPIPS$\downarrow$ \\
\midrule
JPEG & 21.78 & 0.583 & 0.548 & 24.17 & 0.680 & 0.447 & 27.10 & 0.781 & 0.323 \\
\midrule
SD2.1 & 24.11 & 0.669 & 0.371 & 26.43 & 0.755 & 0.277 & 29.20 & 0.837 & 0.194 \\
SD3-medium (ours) & 24.11 & 0.668 & 0.349 & 26.41 & 0.754 & 0.257 & 29.19 & 0.836 & 0.179 \\
\bottomrule
\end{tabular}
\end{table*}

\section{Results on WebP}
\label{app:webp}

To test another codec, we retrain both stages of FDIR on WebP-compressed inputs. The training data, the QF range $[1, 30]$, and all other settings follow those of the JPEG model. Table~\ref{tab:webp} reports results on DIV2K-Val. Compared with the decompressed input, FDIR raises PSNR by 1.30 to 1.34~dB and lowers LPIPS and DISTS at all three quality factors. Compared with the Stage~1 output, FDIR gains 1.47 to 1.98~dB in PSNR, and LPIPS differs by less than 0.005.

\section{Stage~1 Backbone}
\label{app:backbone}

To test another backbone, we replace SD3-medium with Stable Diffusion~2.1 (SD2.1)~\citep{rombachHighResolutionImageSynthesis2022}. We apply LoRA with the same rank to the attention layers of its U-Net and use the QF embedding in place of the text embedding. We train Stage~1 for 230K steps and then retrain FCDR on its output using the same settings as the main model. Table~\ref{tab:backbone} reports results on DIV2K-Val. With SD2.1, PSNR stays within 0.03~dB and SSIM within 0.002 of SD3-medium at all three quality factors. This indicates that FDIR maintains its fidelity with a different backbone. SD3-medium yields lower LPIPS by 0.015 to 0.022, so we use it as the default.

\section{QF Estimation}
\label{app:qf_estimation}

FDIR conditions Stage~1 on the quality factor. A JPEG file stores its quantization tables, and libjpeg derives them from the QF by scaling a standard table. The QF can therefore be read from the file without a trained estimator. We read the luminance table with PIL and match it against the standard table scaled for each QF from 1 to 100.

\textbf{Setup.} We encode LIVE1 (29 images), DIV2K-Val (100 images), and DIV2K-Train (800 images) with OpenCV at QF=1, 5, and 10, and once more at a QF drawn uniformly from 1 to 100 for each image. This gives 3,716 JPEG files. ImageMagick \texttt{identify} gives the same estimates on all files.

\textbf{Results.} Table~\ref{tab:qf_estimation} shows that the estimate equals the encoding QF for every file. One estimate takes 0.0145~ms in a single Python process, measured over 1,000 calls on an in-memory LIVE1 image. This is about 0.03\% of the 49.8~ms FDIR latency (Table~\ref*{tab:latency_comparison}). Calling \texttt{identify} as a separate process takes 6.6~ms per file, including process start-up.

\textbf{Scope.} The test covers files written with the libjpeg standard tables. Files with custom tables, such as some camera outputs, were not tested.

\begin{table}[t]
\centering
\caption{QF estimation accuracy on JPEG files encoded with OpenCV.}
\label{tab:qf_estimation}
\small
\resizebox{\columnwidth}{!}{%
\begin{tabular}{lccc}
\toprule
Dataset & Images & QF=1, 5, 10 & Random QF (1--100) \\
\midrule
LIVE1 & 29 & 100\% (87/87) & 100\% (29/29) \\
DIV2K-Val & 100 & 100\% (300/300) & 100\% (100/100) \\
DIV2K-Train & 800 & 100\% (2,400/2,400) & 100\% (800/800) \\
\bottomrule
\end{tabular}%
}
\end{table}

\section{Model Complexity}
\label{app:model_complexity}

Table~\ref{tab:model_complexity} lists parameters and computation for one $256{\times}256$ input. Stage~1 accounts for most of the computation because of the SD3-medium backbone, while only LoRA and the QF encoder are trained. Stage~2 is trained in full.

\begin{table}[t]
\centering
\caption{Parameters and computation of FDIR for one $256{\times}256$ forward pass. FLOPs: floating-point operations. MACs: multiply-accumulate operations.}
\label{tab:model_complexity}
\small
\setlength{\tabcolsep}{4pt}
\resizebox{\columnwidth}{!}{%
\begin{tabular}{lcccc}
\toprule
 & Trainable (M) & Total (M) & FLOPs (G) & MACs (G) \\
\midrule
Stage 1 & 29.54 & 2,141.69 & 1,372.98 & 686.49 \\
Stage 2 & 43.97 & 43.97 & 135.77 & 67.88 \\
Full pipeline & 73.51 & 2,185.66 & 1,508.75 & 754.38 \\
\bottomrule
\end{tabular}%
}
\end{table}

\section{Additional Visual Comparisons}
\label{app:additional_visualizations}

\begin{figure*}[t]
\centering
\includegraphics[width=\textwidth]{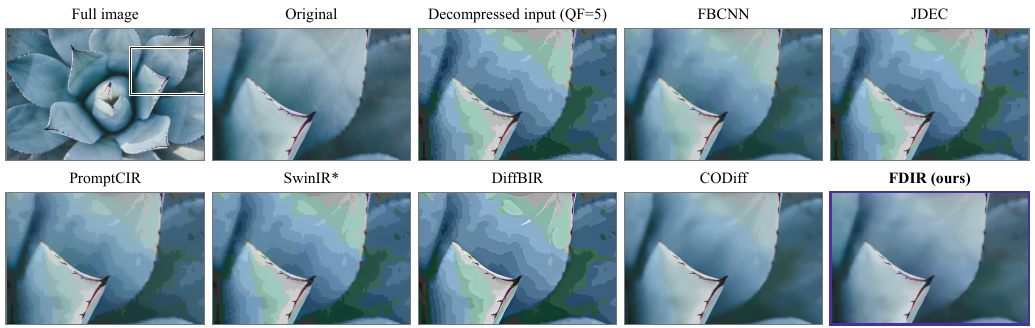}
\caption{Visual comparison at QF=5 on image 0857 from DIV2K-Val. The white box in the full image marks the enlarged region. SwinIR$^*$ uses its QF=10 model. DiffBIR uses tiled inference.}
\label{fig:supp_visual_qf5}
\end{figure*}

\begin{figure*}[t]
\centering
\includegraphics[width=\textwidth]{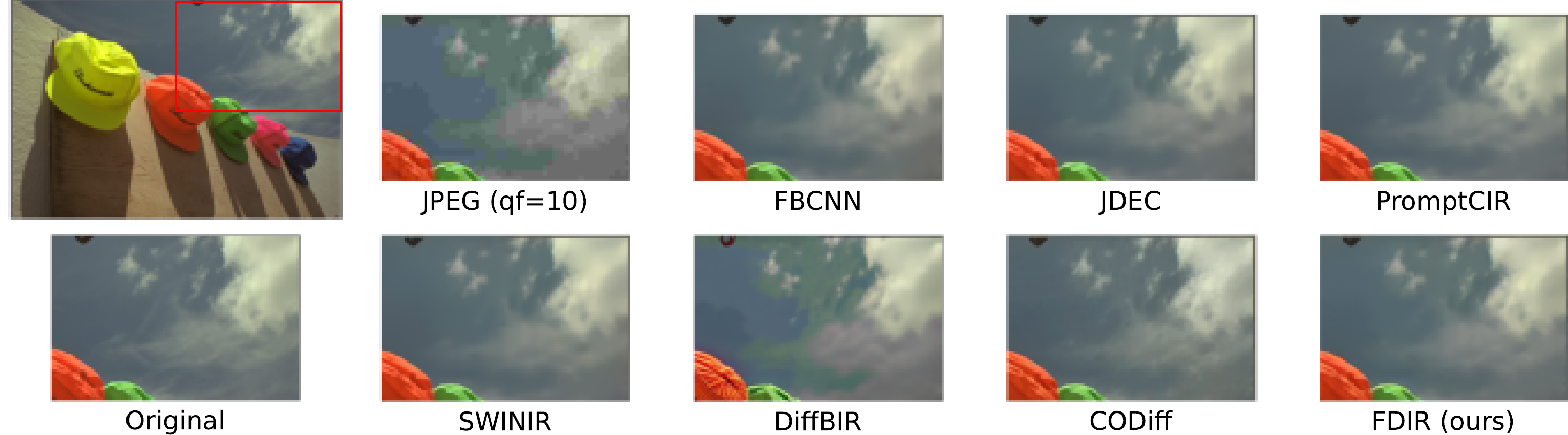}
\caption{Visual comparison at QF=10 on \texttt{caps} from LIVE1. The red box in the original image marks the enlarged region.}
\label{fig:supp_visual_qf10}
\end{figure*}

Further results for all compared methods are shown in Figs.~\ref{fig:supp_visual_qf5} and~\ref{fig:supp_visual_qf10}.

\section{Extended Limitations and Broader Impact}
\label{app:limitations}

\textbf{Codec and QF range.} FDIR is trained and evaluated on JPEG with QF in $[1, 30]$. Other codecs, such as AVIF and HEIC, use different transforms and produce different artifacts. They require retraining, as done for WebP in Sec.~\ref{app:webp}. Higher QF values require a wider conditioning range.

\textbf{Backbone.} Stage~1 uses the 2.1B-parameter SD3-medium backbone. Its prior may carry biases on domains far from natural photographs, such as medical or satellite images. The backbone size also limits deployment on edge devices.

\textbf{Stage~1 errors.} If Stage~1 synthesizes an incorrect structure, Stage~2 may not fully correct it, because Stage~2 uses the prior as guidance.

\textbf{Broader impact.} Stage~1 uses a generative prior that can produce content absent from the original image. Stage~2 anchors the output to the decompressed input, which reduces but does not remove this risk. In medical imaging, forensics, and remote sensing, restored images should be checked against the original data. Restoration can also be misused to alter visual evidence, a risk shared by generative restoration methods.
\end{document}